\documentclass{aip-cp}

\usepackage[numbers]{natbib}
\usepackage{rotating}
\usepackage{graphicx}

\usepackage{graphicx}

\def\nuc#1#2{${}^{#1}$#2}

\def\BBz{$\beta\beta(0\nu)$}

\def\BBt{$\beta\beta(2\nu)$}

\def\BB{$\beta\beta$}

\def\qval{$Q_{\beta\beta}$}                 % The Q-value
\def\be{\begin{equation}}
\def\ee{\end{equation}}

\begin{document}

\title{Solar Neutrinos as Background to Neutrinoless Double-beta Decay Experiments}

\affil[lanl]{Los Alamos National Laboratory, Los Alamos, NM, USA}
\affil[ou]{Research Center for Nuclear Physics and Department of Physics, Osaka University, Ibaraki, Osaka, Japan}

%%%%%%%%%%%%

\author[lanl]{S.R.~Elliott}
\author[ou]{H.~Ejiri}

\maketitle

\begin{abstract}
Solar neutrinos interact within double-beta decay (\BB) detectors and contribute to backgrounds for \BB\ experiments. Background contributions due to charge-current solar neutrino interactions with \BB\ nuclei of  $^{76}$Ge, $^{82}$Se, $^{100}$Mo, $^{130}$Te, $^{136}$Xe, and $^{150}$Nd are evaluated. They are shown to be significant for future high-sensitivity \BB\ experiments that may search for Majorana neutrino masses in the inverted-hierarchy mass region. The impact of solar neutrino backgrounds and their reduction are discussed for future \BB\ experiments.   

\end{abstract}

%%%%%%%%%%%%%%%%%%%%%%%%%%%%%%%%%%%%%%%%%%%%
%% MAINMATTER
%%%%%%%%%%%%%%%%%%%%%%%%%%%%%%%%%%%%%%%%%%%%

\section{Scientific Motivation}
Neutrino-less double beta decay (\BBz) is a unique and realistic probe for studies of neutrino ($\nu$) properties and especially the Majorana mass character of the neutrino and the absolute mass scale. \BB\ studies and $\nu$ masses are discussed in recent reviews and their references \cite{ell04,eji05,avi08,Vergados2012}. 

The rate of \BBz, if it exists, would be extremely small because \BB\ is a second-order weak process that requires lepton number conservation violation and Majorana fermions. For light-neutrino exchange, the rate depends on the effective Majorana mass squared and the typical mass regions to be explored are below 45meV. The \BBz\ half-lives expected for these regions are near or greater than $10^{27}$ years. As a result, the \BBz\  signal rate ($S_{\beta \beta}$) is near or less than a few counts per ton of \BB\ isotope per year (t y).  Accordingly the background rate necessarily has to be around or less than one count per t y.  

Solar-$\nu$s are omnipresent and cannot be shielded, and thus their charged current (CC) and neutral current (NC) interactions are potential background sources for high sensitivity \BB\ experiments as discussed in \cite{Ejiri2014,Ejiri2017,deBarros2011} and references therein. In fact, it has been shown that solar-$\nu $ CC interactions with \BB\ isotopes like $^{100}$Mo~\cite{eji00}, $^{116}$Cd~\cite{Zuber2003} and $^{150}$Nd~\cite{Zuber2012} can be used for real-time studies of the low-energy solar-$\nu$s. 

The \BB\ isotopes most often used or considered for high-sensitivity experiments are $^{76}$Ge, $^{82}$Se, $^{100}$Mo, $^{130}$Te, $^{136}$Xe and $^{150}$Nd.  Here, we classify them into two groups. Group A consisting of $^{82}$Se, $^{100}$Mo  and $^{150}$Nd have a large solar-$\nu $ CC rate, whereas Group B consisting of $^{76}$Ge, $^{130}$Te and $^{136}$Xe have a rather small solar-$\nu$ CC rate.  Solar-$\nu$ interactions with atomic electrons in \BB\ isotopes and liquid scintillators used for \BB\ experiments were considered in Refs.~\cite{ell04,bar11,deBarros2014,Ejiri2016}. 
 
The present paper aims to summarize the background contributions for these 6 nuclei and discuss the impact on high sensitivity \BB\ experiments using them.  The mechanics of these calculations are identical to those in Ref.~\cite{Ejiri2014,Ejiri2017}, and we do not repeat all the details here. 

\section{Solar Neutrino Backgrounds}

\begin{figure}[htb]
\caption{Left: Schematic diagrams of the \BB\ of $^{Z-1}$A to $^{Z+1}$A, the solar-$\nu $ CC interaction on $^{Z-1}$A and the electron ($\gamma $) 
decay to $^{Z}$A,  and  the single $\beta $/$\gamma $ decays of $^{Z}$A to $^{Z+1}$A. $Q_{\beta \beta }$, $Q_{\beta}$ , $Q_{\nu}$, and $Q_{e}$  are the $Q$
values for ${\beta \beta }$, $\beta $, $\nu $ CC and electron capture (EC), respectively. Right: Solar-$\nu $ capture rates in units of SNU for current \BB\ nuclei  are plotted against the neutrino CC $Q_{\nu}$ value for the lowest 1$^+$ state.  $S_{t}$ is the total capture rate. Group A nuclei have a large $S_{t}$. Group B nuclei have a small  $S_{t}$.
\label{fig:solar}}
%\begin{center}
\includegraphics[width=7.5 cm]{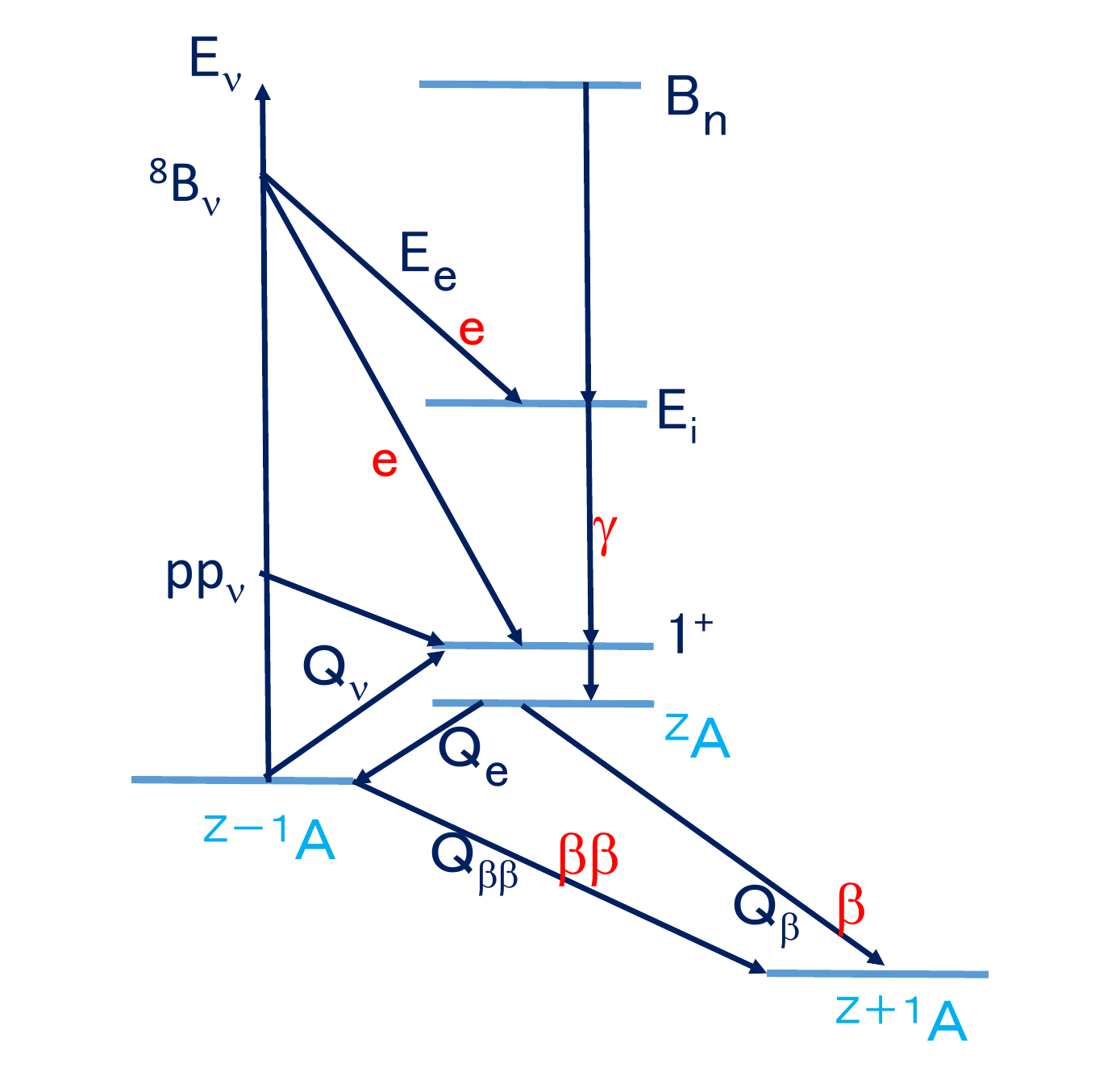}
\includegraphics[width=7.5 cm]{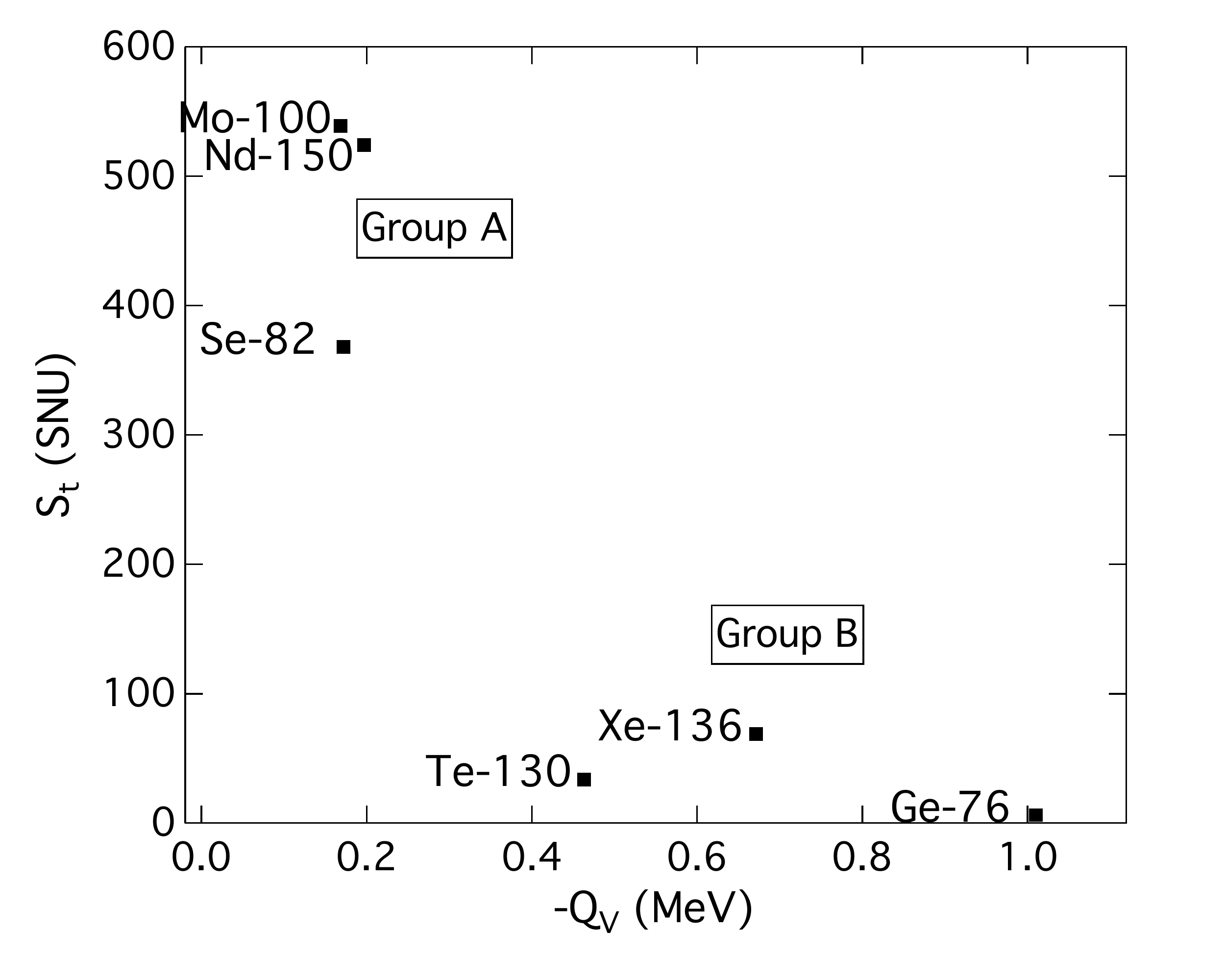}
%\end{center}
\end{figure}

The process of \BB\ decay from $^{Z-1}$A to  $^{Z+1}$A via the intermediate nucleus $^{Z}$A is shown in Eqn.~\ref{eqn:Processes}. Solar $\nu$s can produce background to this signal primarily through CC interactions with \BB\ nuclei. The CC interaction produces background in two ways. First the CC interaction itself can produce a signal ($B_{CC}$) given by the promptly emitted e$^-$ and, if the resulting nucleus is in an excited state, a number of $\gamma$ rays may be emitted as the nucleus relaxes to its ground state. Second, the resulting nucleus $^{Z}$A can then $\beta^-$ decay to $^{Z+1}$A by emitting a single $\beta^-$ ray and possibly also $\gamma $ ray(s) if the residual state is an excited state ($B_{SB}$). The interaction and decay schemes are shown in Fig.~\ref{fig:solar}. The 3 processes are expressed as,
\begin{eqnarray}
\beta\beta:~~  ^{Z-1}A 		&   \rightarrow	& ^{Z+1}A+ \beta^- +\beta^- +Q_{\beta \beta } \nonumber	\\
CC:~~		 ^{Z-1}A+\nu	&  \rightarrow 	& ^{Z}A +e^- +\gamma(s) +Q_{\nu}\nonumber	\\
SB:~~		 ^{Z}A		&  \rightarrow 	& ^{Z+1}A  + \beta^- +\gamma(s) +Q_{\beta},
\label{eqn:Processes}
\end{eqnarray}
where \BB, CC, and SB denote the double beta decay, the solar-$\nu$ CC interaction, and the single beta decay processes, respectively. The $Q$ values for each are given as $Q_{\beta \beta }$, $Q_{\nu}$, and $Q_{\beta }$ respectively, as shown in Fig.~\ref{fig:solar}. The potential background from CC at the \qval\ is much smaller than that from SB, and hence we only discuss the SB case.

We consider \BB\ detectors where the sum energy of the $\beta $ and $\gamma $ rays is measured. The \BBz\ signature is a peak within the region of interest (ROI) at the \BB\ Q value ($E=Q_{\beta \beta }$).  In contrast, the sum of the electron energy ($E_e$) and any successive $\gamma $ ray energy ($E_{\gamma }$) is a continuum spectrum for both the CC and SB processes. The total observed energy from the CC and SB processes can exceed \qval\ resulting in a background that populates the \BBz\ region of interest. The backgrounds of interest are estimated by their yields within the ROI and therefore are sensitive to the detector's energy resolution.  The energy width of the ROI is given by the FWHM resolution ($\Delta E$) at the energy of the Q value. This ratio, defined as $\delta \def \Delta E/Q_{\beta \beta }$, is the fractional energy resolution at $E=Q_{\beta \beta }$. Certainly the kinematics and event topologies for the backgrounds discussed here may permit experimental techniques for rejection that go beyond just a sum energy cut. We cannot anticipate all possible rejection techniques, but instead estimate rates so that future efforts may better assess these backgrounds for a given experimental configuration.  

The solar-$\nu$ capture rates for individual neutrino sources are evaluated by using the neutrino responses ($B(GT), B(F)$) given by recent charge exchange reactions and the neutrino fluxes from BP05(OP)~\cite{bah05}. We used the charge-exchange-measurement data for \nuc{76}{Ge}~\cite{Thies2012}, \nuc{82}{Se}~\cite{Frekers2016}, \nuc{100}{Mo}~\cite{Thies2012a}, \nuc{130}{Te}~\cite{Puppe2012}, \nuc{136}{Xe}~\cite{Frekers2013} and \nuc{150}{Nd}~\cite{Guess2011} to obtain the $B(GT)_k$ and $B(F)$ for the IAS for the indicated isotopes. The calculations were done as in Ref.~\cite{Ejiri2014,Ejiri2017} including the treatment of $\nu$ oscillations. The \BB, solar-$\nu$ interaction, and single $\beta$ $Q$ values and the solar-$\nu$ capture rates are shown Table~\ref{tab:SolarRates}.
 
\begin{table}[htb]
\small
\caption{Key nuclear physics data for the isotopes under consideration.  $Q_{\beta \beta}$ is the \BB\ $Q$ value, $Q_{\nu}$ is the $\nu$-CC $Q$ value for the lowest 1$^+$ state, $Q_{\beta}$ is the single $\beta$ Q value.
\label{tab:Isotopes}}
\centering
\vspace{0.5cm}
\begin{tabular}{lcccc}
\hline
Isotope 	& \BBt\ $\tau_{1/2}$				& $Q_{\beta \beta }$ 	& $Q_{\nu}$ 	& $Q_{\beta}$  \\
		& years 						& MeV 			& MeV 		&  MeV 		 \\
\hline\hline
$^{76}$Ge & $1.93\times10^{21}$~\cite{Agostini2015} 	& 2.039 		&~~ -1.010 		& ~~2.962 	 \\
$^{82}$Se & $9.2\times10^{19}$~\cite{Barabash2015}	& 2.992 			&~~ -0.172		& ~~3.093		 \\
$^{100}$Mo & $7.1\times10^{18}$~\cite{Barabash2015}	& 3.034 		&~~ -0.168 		& ~~3.202 		\\
$^{130}$Te & $6.9\times10^{20}$~\cite{Barabash2015}	& 2.528 		& ~~-0.463		& ~~2.949 		 \\
$^{136}$Xe &$2.19\times10^{21}$~\cite{Barabash2015}	& 2.468 		& ~~-0.671 		& ~~2.548 		\\
$^{150}$Nd & $8.2\times10^{18}$~\cite{Barabash2015}	& 3.368 		&~~ -0.197 		& ~~3.454 		 \\
\hline\hline
\end{tabular}
\end{table}

\begin{table}[htb]
\caption{\BB, CC, and SB $Q$ values in units of MeV and solar-$\nu $ capture rates in units of SNU for selected \BB\ nuclei including the effect of oscillations.  Column 3 gives $S_t$ for no oscillations., $S_{B}$ is the $^8$B-$\nu$ capture rate, and $S_{t}$ is the total solar-$\nu$ capture rate. The background rates for $\beta$ decay ($B_{SB}$) and \BBt\ ($B_{2\nu}$) are calculated for $\delta = 0.02$. 
\label{tab:SolarRates}}
\centering
\vspace{0.5cm}
\begin{tabular}{lcccccc}
\hline
Isotope 	& $S_{pp}$ 	& $S_{B}$  & $S_t$ no osc. 	&$S_t$   		& $B_{SB}$ 	& $B_{2\nu}$ \\
		&  (SNU)  		& (SNU) 	& (SNU) 			& (SNU) 		& events/ t y 	& events /t y \\
\hline\hline
$^{76}$Ge & 0 			&  \phantom{0}5.0  & \phantom{0}15.7  	& \phantom{00}6.3 & 0.03		&0.005 \\
$^{82}$Se & 257 		& 10.0	& 672  			&  368		& 4.42		&0.15\phantom{0} \\
$^{100}$Mo & 391 		&\phantom{0}6.0 & 975 		&  539		&  0.11		&1.56\phantom{0}\\
$^{130}$Te & 0			&\phantom{0}6.1   &  \phantom{0}67.7 		&  \phantom{0}33.7 		&0.48		&0.01\phantom{0} \\
$^{136}$Xe & 0			&\phantom{0}9.8& 136 		&   \phantom{0}68.8		& 0.55		&0.003\\
$^{150}$Nd & 352 		&15.5	&  961  			&  524		& 0.12		&1.00\phantom{0}  \\
\hline\hline
\end{tabular}
\end{table}

\section{Discussion}
Total solar-$\nu$ capture rates $S_t$ for Group A and B nuclei are plotted against the neutrino CC $Q_{\nu}$ value in Fig.~\ref{fig:solar}. The Group A nuclei of $^{82}$Se, $^{100}$Mo  and $^{150}$Nd with small negative $Q_{\nu}$ values around -170 keV, have large solar-$\nu$ capture rates because they are strongly excited by the pp neutrinos. On the other hand Group B nuclei of $^{76}$Ge, $^{130}$Te and $^{136}$Xe have rather small solar-$\nu$ CC rates because the threshold energy (-$Q_{\nu}$) is large enough that the pp neutrinos can not excite them. The pp neutrino contributions to the total capture rates for the Group A nuclei  are around 60\% of the total, while the $^7$Be and $^8$B neutrino capture rate are around 30\% and 3\%, respectively.

The probability that an event falls within the peak region for \BBz ($\epsilon _{SB}$) is evaluated as a function of $\delta$ for simple calorimetric detectors. The background rates are approximately proportional to the resolution, i.e. the width of the ROI. $B_{SB}$ for a fractional resolution of $\delta =0.02$ are given in Table~\ref{tab:SolarRates}. 

\begin{table}
\caption{A qualitative summary of the potential cuts available to reject background from the CC and SB processes. There are experimental proposals for \nuc{82}{Se} and \nuc{100}{Mo} using bolometers, which if successful, will have good energy resolution. These were listed as {\em maybe} in the resolution column.
\label{tab:BGRejection}}
\centering
\vspace{0.5cm}
\begin{tabular}{lcccc}
\hline
Isotope 				& CC Rate 	& CC Tag 		& Resolution 	&MSE     \\
\hline\hline
$^{76}$Ge 	 		&   low		&	no		&	yes		&some	\\
$^{82}$Se 	 		& 	high		&	no		&  	maybe	& yes  \\
$^{100}$Mo 	 		&	high		&	yes		&	maybe	& no	\\
$^{130}$Te 			&	low		&	no		&	yes		& yes  \\
$^{136}$Xe 			&	medium	&	no		&			&yes		\\
$^{150}$Nd 	 		& 	high		&	no		&			& yes\\
\hline\hline
\end{tabular}
\end{table}

\nuc{76}{Ge} has a low CC rate due to the high $Q_{\nu}$. The intermediate nucleus, \nuc{76}{As} has a 1-day half-life making a tag by delayed coincidence difficult for the CC interaction, and about half of the decay is pure $\beta$ decay permitting multiply energy site (MSE) deposit cuts only useful in about half the decays. Hence one relies on the low rate and good energy resolution of Ge detectors to mitigate this background. \nuc{82}{Se} has a large CC rate and \nuc{82}{Br} has a 35 d half-life making the CC tag difficult. The decay of \nuc{82}{Br} however is always to an excited state permitting MSE cuts. \nuc{100}{Mo} has a large CC rate, but \nuc{100}{Tc} has a short half-life of 16 s. This permits a CC tag. A large fraction of the \nuc{100}{Tc} decays are to the ground state of \nuc{100}{Ru} making a MSE cut hard. \nuc{130}{Te} has a low CC rate and \nuc{130}{I} has a 12 h half-life making a CC tag difficult. The decays are entirely to an excited state permitting a good MSE cut.  \nuc{136}{Xe} has a fairly low CC rate, but \nuc{136}{Cs} has a 13 d half-life making a CC tag difficult. The decay is always to an excited state permitting a good MSE cut. \nuc{150}{Nd} has a high CC rate and \nuc{150}{Pm} has a 3 h half-life making a CC tag difficult. Most of the decay is to an excited state leading to a good MSE cut. These issues are summarized in Table~\ref{tab:BGRejection}. We note that the measurement of individual $\beta$-ray tracks would help separate \BB\ signals from SB background, but building such detectors with a ton-scale quantity of isotope will be a challenge.

\bibliographystyle{aipnum-cp.bst}
\bibliography{DoubleBetaDecay.bbl}

\end{document}